\documentclass[preprint,preprintnumbers,amsmath,amssymb]{revtex4}

\usepackage{color}
\usepackage{graphicx}
\usepackage{bm}

\begin{document}

\includegraphics*[viewport=90 50 600 730, page=1]{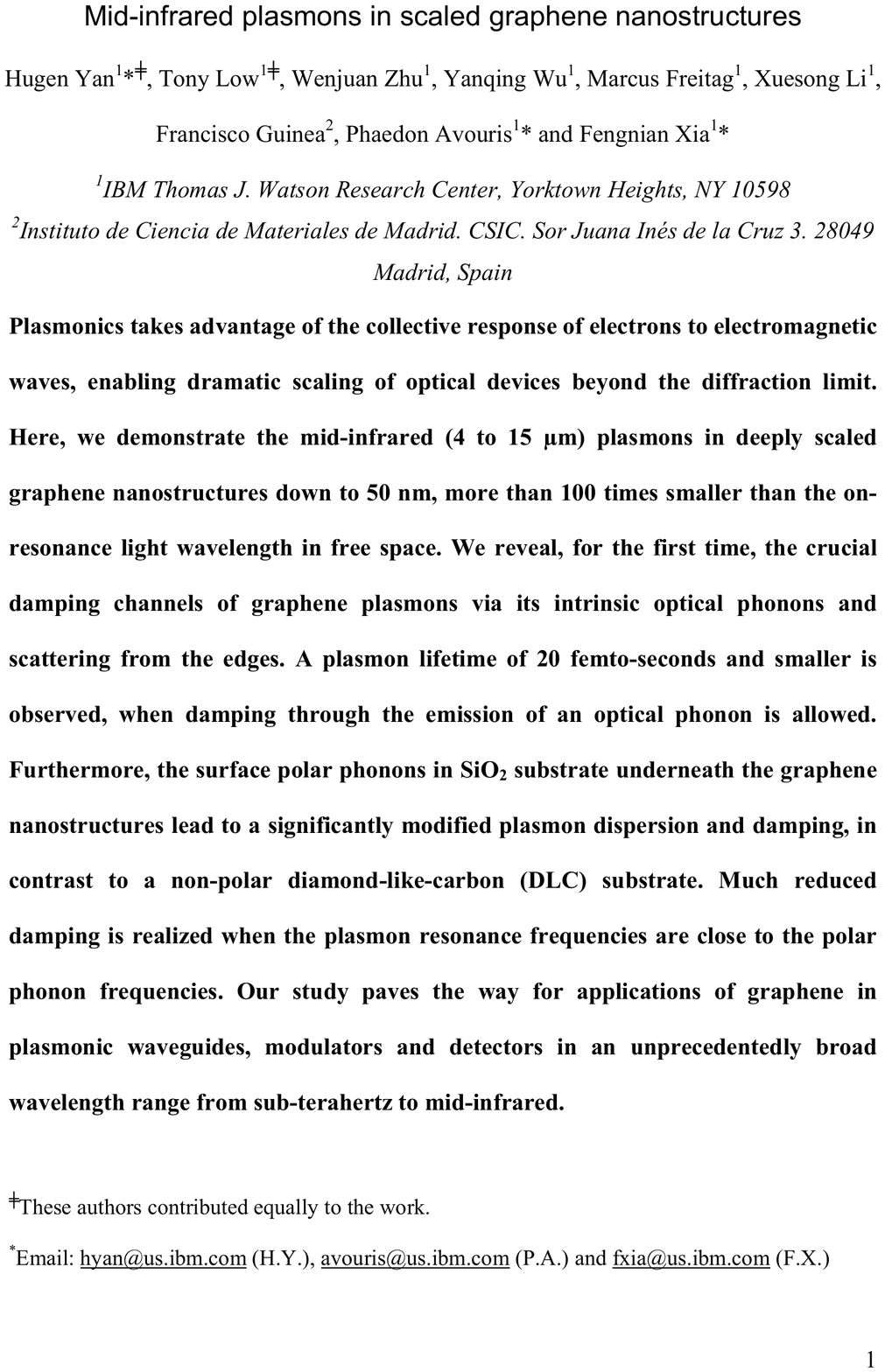}
\includegraphics*[viewport=80 50 600 730, page=2]{main.pdf}
\includegraphics*[viewport=80 50 600 730, page=3]{main.pdf}
\includegraphics*[viewport=80 50 600 730, page=4]{main.pdf}
\includegraphics*[viewport=80 50 600 730, page=5]{main.pdf}
\includegraphics*[viewport=80 50 600 730, page=6]{main.pdf}
\includegraphics*[viewport=80 50 600 730, page=7]{main.pdf}
\includegraphics*[viewport=80 50 600 730, page=8]{main.pdf}
\includegraphics*[viewport=80 50 600 730, page=9]{main.pdf}
\includegraphics*[viewport=80 50 600 730, page=10]{main.pdf}
\includegraphics*[viewport=80 50 600 730, page=11]{main.pdf}
\includegraphics*[viewport=80 50 600 730, page=12]{main.pdf}
\includegraphics*[viewport=80 50 600 730, page=13]{main.pdf}
\includegraphics*[viewport=80 50 600 730, page=14]{main.pdf}
\includegraphics*[viewport=80 50 600 730, page=15]{main.pdf}
\includegraphics*[viewport=80 50 600 730, page=16]{main.pdf}
\includegraphics*[viewport=80 50 600 730, page=17]{main.pdf}
\includegraphics*[viewport=80 50 600 730, page=18]{main.pdf}
\includegraphics*[viewport=80 50 600 730, page=19]{main.pdf}
\includegraphics*[viewport=80 50 600 730, page=20]{main.pdf}
\includegraphics*[viewport=80 50 600 730, page=21]{main.pdf}
\includegraphics*[viewport=80 50 600 730, page=22]{main.pdf}
\includegraphics*[viewport=80 50 600 730, page=23]{main.pdf}
\includegraphics*[viewport=80 50 600 730, page=24]{main.pdf}
\includegraphics*[viewport=80 50 600 730, page=25]{main.pdf}

\title{Supplementary Information:\\ 
Mid-infrared plasmons in scaled
graphene nanostructures}

\author{Hugen Yan$^{1\dagger}$, Tony Low$^{1\dagger}$, Wenjuan Zhu$^1$,
Yanqing Wu$^1$, Marcus Freitag$^1$, Xuesong Li$^1$,
Francisco Guinea$^2$, Phaedon Avouris$^1$ and Fengnian Xia$^1$}
\affiliation{$^1$ IBM T.J. Watson Research Center, Yorktown Heights, NY 10598, USA\\
$^2$ Instituto de Ciencia de Materiales de Madrid. CSIC. Sor Juana In\'es de la Cruz 3. 28049 Madrid, Spain\\
\newline
$^\dagger$\small
These authors contributed equally to this work\normalsize
}
%
\maketitle

\section{Calculating the RPA dielectric function and loss function}

This section describes the modeling of the loss function in graphene on SiO$_2$
as shown in the intensity plot of Fig.\,3b in the main manuscript.

\subsection{Intrinsic phonons}

The only intrinsic phonons with momenta and energies similar to the graphene plasmons in our experiment are the
longitudinal/transverse optical (LO/TO) phonons near the $\Gamma$ point,
with energies $\hbar\omega_{op} \approx 0.2\,$eV.
On symmetry grounds\cite{M07}, their coupling to electrons can be written as\cite{A06_OP,IA06_OP}
\begin{align}
{\cal H}_{e\mbox{-}op}(\bold{r}) &= g_{0} \left( \begin{array}{cc} 0 &u_y(\bold{r}) + i u_x(\bold{r})
\\ u_y(\bold{r}) - i u_x(\bold{r}) &0 \end{array} \right) =
\frac{g_{0}}{v_F} \hat{\bold{j}}(\bold{r})\times \bold{u}(\bold{r})
\end{align}
where $v_F$ is the Fermi velocity and $g_{0}$ the coupling constant. The coupling constant can be estimated from the change with bond length of the hopping between nearest neighbor carbon $\pi$ orbitals\cite{NG07,B08}, $g_0 \approx \partial t / \partial l$.
The electronic Hamiltonian is described within each valley (and spin) in terms of the amplitudes on A/B sublattices,
$\hat{\bold{j}}(\bold{r})$ is the single-particle current operator
and $\bold{u}(\bold{r})$ is the relative displacement of the two sublattices.
Their representation in terms of electron and phonon ladder operators, i.e. $\hat{a}_{\bold{k}}$ and $\hat{b}_{\bold{q}}$ respectively, are given by,
\begin{align}
\hat{\bold{j}}(\bold{r}) =\frac{1}{A} \sum_{\bold{kq}} v_F \hat{a}^{\dagger}_{\bold{k}}\hat{\bold{\sigma}} \hat{a}_{\bold{k+q}} e^{i\bold{q}\cdot\bold{r}}
\equiv \frac{1}{A} \sum_{\bold{q}} \hat{\bold{j}}_{\bold{q}}e^{i\bold{q}\cdot\bold{r}}\\
\bold{u}(\bold{r}) = \sqrt{\frac{\hbar}{2\rho_m A\omega_{op}}}\sum_{\bold{q}\lambda} (\hat{b}_{\bold{q}\lambda}+\hat{b}_{\mbox{-}\bold{q}\lambda}^{\dagger})
\bold{e}_{\bold{q}\lambda}  e^{i\bold{q}\cdot\bold{r}}
\end{align}
where $\hat{\bold{\sigma}}$ are the Pauli spin matrices and $\bold{e}_{\bold{q}\lambda}$
are the polarization vectors. $\lambda$ denotes the phonon modes, $\rho_m$
is the mass density of graphene and $A$ is its area.
Using standard perturbation techniques, the effective electron-electron interaction mediated by
optical phonons can be written as,
\begin{align}
{\cal V}^{op}_{el\mbox{-}el}= \frac{1}{A^2}\sum_{\bold{q}\lambda} \frac{1}{v_F^2}\left|M_{op}\right|^2{\cal D}^0_{\lambda}(\omega)
 \hat{\bold{j}}_{\bold{q}}\cdot\hat{\bold{j}}_{\mbox{-}\bold{q}}
 \equiv \frac{1}{A^2}\sum_{\bold{q}\lambda} v_{op,\lambda}
 \hat{\bold{j}}_{\bold{q}}\cdot\hat{\bold{j}}_{\mbox{-}\bold{q}}
\end{align}
where the scattering matrix elements and the free phonon Green's function are
\begin{align}
|M_{op}|^2 =  \frac{\hbar g^2_0}{2\rho_m\omega_{op}} \mbox{\,\,\,\,\, , \,\,\,\,\,\,}
{\cal D}^0_{\lambda}(\omega)=\frac{2\omega_{op}}{\hbar((\omega+i\hbar/\tau_{op})^2-\omega^2_{op})}
\end{align}
where $\tau_{op}$ phenomenologically describes the phonon lifetime.

\subsection{Surface polar phonons}

Polar substrates, such as SiO$_2$ and BN have optical piezoelectric modes at energies $\hbar\omega_{sp}$. These modes induce electric fields which couple to the carriers in graphene\cite{FG08,SSG12}.
At long wavelengths, the effect of these can be described in terms of the dielectric function of the substrate,
\begin{align}
{\cal H}_{e\mbox{-}sp} = \frac{1}{A}\sum_{\bold{kq}}M_{sp} \hat{a}^{\dagger}_{\bold{k+q}}\hat{a}_{\bold{k}}
(\hat{b}_{\bold{q}\lambda}+\hat{b}_{\mbox{-}\bold{q}\lambda}^{\dagger})
\end{align}
with the scattering matrix elements defined as,
\begin{align}
|M_{sp}|^2 = \frac{\pi e^2}{\epsilon_0}\frac{e^{-2qz_0}}{q}{\cal F}^2 \mbox{\,\,\,\,\, , \,\,\,\,\,\,}
{\cal F}^2 = \frac{\hbar\omega_{sp}}{2\pi}\left(\frac{1}{\epsilon_{high}+\epsilon_{env}}-
\frac{1}{\epsilon_{low}+\epsilon_{env}}\right)
\end{align}
where $z_0$ is the graphene-substrate separation, ${\cal F}^2$
describes the Fr$\ddot{o}$hlich coupling strength,
$\epsilon_{low}$ ($\epsilon_{high}$) are the low (high)
frequency dielectric constant of the dielectric
and $\epsilon_{env}$ is that of the environment.
The effective electron-electron interaction mediated by
surface optical phonons calculated from standard perturbation techniques yields\cite{BF04},
\begin{align}
{\cal V}^{sp}_{el\mbox{-}el}= \frac{1}{A^2}\sum_{\bold{q}\lambda} \left|M_{sp}\right|^2{\cal D}^0_{\lambda}(\omega)
 \hat{\bold{\rho}}_{\bold{q}}\hat{\bold{\rho}}_{\mbox{-}\bold{q}}
\equiv \frac{1}{A^2}\sum_{\bold{q}\lambda} v_{sp,\lambda}
 \hat{\bold{\rho}}_{\bold{q}}\hat{\bold{\rho}}_{\mbox{-}\bold{q}}
\end{align}
where $\hat{\bold{\rho}}_{\bold{q}}\equiv \sum_{\bold{k}} \hat{a}^{\dagger}_{\bold{k}}\hat{a}_{\bold{k+q}}$
and ${\cal D}^0_{\lambda}(\omega)$ contains also a phenomenological phonon lifetime
of $\tau_{sp}$.

\subsection{Dielectric response}

The plasmon response of graphene begins with finding the dielectric function.
A satisfactory approximation can be obtained by adding the separate contributions \emph{independently}.
An effective interaction between electrons is given by the
sum of the direct Coulomb interaction $v_c(q)=e^2/2q\epsilon_0$
and the two electrons interaction mediated by surface phonon $v_{sp,\lambda}(q,\omega)$.
The RPA expansion of the dielectric function, $\epsilon^{rpa}_T(q,\omega)$, can be expressed with this effective interaction\cite{mahan00,HSDS10}
\begin{align}
v_{eff}(q,\omega)=\frac{v_c(q)}{\epsilon_T^{rpa}(q,\omega)}=\frac{v_c(q)+\sum_{\lambda}v_{sp,\lambda}}{1-[v_c(q)+\sum_{\lambda}v_{sp,\lambda}]
\Pi_{\rho,\rho}^0(q,\omega)}
\end{align}
where $\Pi_{\rho,\rho}^0(q,\omega)$ is the non-interacting part (i.e. the pair bubble diagram) of
the charge-charge correlation function given by a modified Lindhard function\cite{WSSG06,HSS07},
\begin{align}
\Pi_{\rho,\rho}^0(q,\omega)=-\frac{g_s}{(2\pi)^2}\sum_{nn'}\int_{\mbox{BZ}}d\bold{k}
\frac{n_F(\xi_{\bold{k}})-n_F(\xi_{\bold{k+q}})}{\xi_{\bold{k}}-\xi_{\bold{k+q}}+\hbar\omega+i\hbar/\tau_e}F_{nn'}(\bold{k},\bold{q})
\end{align}
where $n_F(\xi_{\bold{k}})$ is the Fermi-Dirac distribution function,
$F_{nn'}(\bold{k},\bold{q})$ is the band overlap function of Dirac spectrum
and $\tau_e$ is the lifetime of electrons.
While the polar surface phonons couple to the charge density operator,
the intrinsic optical phonon couple instead to the current operator.
Its contribution to the dielectric function is given by $v_{op}(q,\omega)\Pi^0_{j,j}(q,\omega)$,
where $\Pi^0_{j,j}(q,\omega)$ is the current-current correlation function.
We note that from the usual charge continuity equation, $i\partial_t\hat{\rho}_{\bold{q}}=\bold{q}\cdot\hat{\bold{j}}_{\bold{q}}$,
it follows that, 
\begin{align}
q^2\Pi_{j,j}(q,\omega)  = \omega^2\Pi_{\rho,\rho}(q,\omega) - v_F
\left\langle \left[\bold{q}\cdot\hat{\bold{j}}_{\bold{q}},\hat{\rho}_{-\bold{q}}\right]\right\rangle
\label{chargecurr}
\end{align}
where the second term in Eq.\,\ref{chargecurr} is purely real
and $\propto q^2$ as calculated in Ref.\,\cite{SNCN08}. 
The imaginary part of $\Pi_{j,j}(q,\omega)$ can be obtained just from
$\Im[\tfrac{\omega^2}{q^2}\times\Pi_{\rho,\rho}(q,\omega)]$.
Collective modes with self consistent oscillations of the carrier charge can be obtained from the zeros of the full dielectric function
\begin{align}
\epsilon_T^{rpa}(q,\omega)=\epsilon_{env}-v_{c}\Pi^0_{\rho,\rho}(q,\omega)-
\epsilon_{env} \sum_{\lambda}v_{sp,\lambda} \Pi^0_{\rho,\rho}(q,\omega)
-\epsilon_{env} v_{op}\Pi^0_{j,j}(q,\omega)
\label{erpatotal}
\end{align}
where $\epsilon_{env}$ is the dielectric constant of graphene's environment.

\subsection{Loss function}

Our spectroscopy experiments measure the extinction spectra
defined as $Z\equiv-\delta T/T_0$ with $\delta T = T-T_0$,
where $T$ ($T_0$) is the measured transmission with (without) plasmon excitations.
In the experiment, a superlattice of graphene
ribbons of width $W$ defines the momentum i.e. $q=\pi/(W-W_0)$.
$W_0$ accounts for the difference between physical and electrical device's width.
Varying the frequency of the incident light excitation, $\omega$, polarized
perpendicularly (and parallely) to the ribbon, allows one to quantify the extinction spectra $Z(q,\omega)$,
as first demonstrated in Ref.\,\cite{JGHGM11}.
Resonance peaks in $Z(q,\omega)$ corresponds to enhanced
optical absorption by graphene originating from plasmon oscillations\cite{JGHGM11,KCG11,NGGM12} and
can best be described by,
\begin{align}
Z(q,\omega)  \sim -\Im\left[\frac{1}{\epsilon_T^{rpa}}\right]
\end{align}
where the latter is known as the loss function,
which describes the ability of the system to dissipate energy via plasmon excitations
and can be calculated from Eq.\,\ref{erpatotal}.\\

Using the above theory, we plot the loss function in graphene on SiO$_2$
as shown in Fig.\,3b of the main manuscript.
The calculations include interactions with the intrinsic and SiO$_2$ substrate phonons.
Graphene doping is assumed to be $E_f=-0.43\,$eV and an effective $\epsilon_{env}=1.5$.
The frequencies of the various phonon modes are assumed to be at
$\omega_{op}=1580\,$cm$^{-1}$, $\omega_{sp1}=806\,$cm$^{-1}$ and $\omega_{sp2}=1168\,$cm$^{-1}$.
The damping time used in those plots are
$\tau_e=0.1\,$ps, $\tau_{op}=70\,$fs and $\tau_{sp}=1\,$ps.
The coupling parameters used are $g_0=7.7\,$eV$\AA^{-1}$,
${\cal F}^2_{sp1}=0.2\,$meV and ${\cal F}^2_{sp2}=2\,$meV.
Note that another substrate phonon at $\omega_{op}=460\,$cm$^{-1}$
was not included in the calculation,
given that our experiment data are far above that frequency.

\section{Calculating lifetimes of the plasmon and coupled plasmon-phonon modes}

This section describes the modeling of the plasmon lifetime in graphene on DLC and SiO$_2$
as shown in Fig.\,4a and 4c of the main manuscript.\\

In the above previous analysis, the damping mechanisms for the plasmons
are not discussed.
Exchange of energy and momentum during scattering of plasmons
can bring it into the Landau damping regime, leading to finite damping.
In fact, when the plasmon energy exceeds the optical phonon energy,
it can decay into a phonon together with an electron-hole pair, in such a way that the total momentum is conserved.
In a phenomenological way, this decay can be accounted for through the single particle excitations,
which have a finite lifetime $\tau_e$, when their energies exceed the optical phonon energy\cite{JBS09}, for example.
Damping related to scattering with the ribbon's edges and
a background damping due to impurities in the bulk can also be incorporated in $\tau_e$.
Finite phonon lifetime, $\tau_{sp}$, can also influence to plasmon damping
in the coupled plasmon-phonon modes.
Below, we present our description of plasmon damping in the presence
and absence of coupling with the surface phonon modes.\\

We are interested in the regime where $\omega>v_F q$ and $E_f\gg\hbar\omega$. In this limit,
\begin{align}
\Pi^0_{\rho,\rho}(q,\omega)\approx \frac{E_f q^2}{\pi\hbar^2(\omega+i\delta_{e})^2}
\end{align}
where $\delta_{e}$ is
the single-particle related damping in graphene
defined as $\delta_{e}\equiv 1/\tau_e$.
In the absence of substrate phonon interactions, such as
the case of graphene on a DLC substrate,
the plasmon frequency is simply
$\omega=\omega_{pl}+i\delta_{e}$, where
$\omega_{pl}^2=q |E_f| e^2/2\pi\hbar^2\epsilon_0\epsilon_{env}$
and $\delta_{e}$ also corresponds to plasmon damping.
Note that in the regime we are considering i.e. $\omega>v_F q$ and $E_f\gg\hbar\omega$,
Landau damping is excluded.
Guided by experiments, the plasmon hybridizes strongly with one of the surface
phonon modes with $\omega_{sp}\approx 0.145\,$eV on SiO$_2$ substrate. For $\omega>\omega_{sp}$,
we can write a simpler dielectric function,
\begin{align}
\epsilon_{T}^{rpa}\approx \epsilon_{env}\left[1-\frac{\omega_{pl}(q)^2}{(\omega+i\delta_{e})^2}-
\frac{\tilde{\omega}_{sp}^2}{(\omega+i\delta_{sp})^2-\omega_{sp}^2+\tilde{\omega}_{sp}^2}\right]  \,\,\,\,\,\,\,\,\,\mbox{,}\,\,\,\,\,\,\,\,\,
\tilde{\omega}_{sp}\equiv \sqrt{\frac{4\pi}{\hbar}\omega_{sp}{\cal F}^2}
\label{eprpasim}
\end{align}
where $\delta_{sp}$ is
the surface phonons damping rate defined as $\delta_{sp}\equiv 1/\tau_{sp}$.
The frequencies of the coupled plasmon-phonon modes can be
obtained by setting $\epsilon_{T}^{rpa}=0$ i.e.,
\begin{align}
\omega^4+i2\omega^3(\delta_{sp}+\delta_{e})-\omega^2(\omega_{sp}^2+\omega_{pl}(q)^2)
-i2\omega(\omega_{sp}^2\delta_{e}+\omega_{pl}(q)^2\delta_{sp})
+\omega_{pl}(q)^2(\omega_{sp}^2-\tilde{\omega}_{sp}^2)=0
\label{plasphomodes}
\end{align}
which can be solved numerically. In the limit where
$\delta_{e}=\delta_{sp}=0$, it reduces to a simple biquadratic
equation with coupled plasmon-phonon modes solutions given by,
\begin{align}
\omega_{\pm}^2=\frac{\omega_{pl}^2+\omega_{sp}^2}{2}
\pm \frac{\sqrt{(\omega_{pl}^2+\omega_{sp}^2)^2-4\omega_{pl}^2(\omega_{sp}^2-\tilde{\omega}_{sp}^2)}}{2}
\end{align}
In the general case where $\delta_{e}=\delta_{ph}\neq 0$, we solve for the coupled plasmon-phonon modes
via Eq.\,\ref{plasphomodes} numerically. However, in the $q=0$ limit,
it can be shown by setting $\epsilon_{T}^{rpa}=0$ in
Eq.\,\ref{eprpasim} that $\omega=\omega_{sp}-i\delta_{sp}$.
Therefore, the lifetime 
of the plasmon with frequency
in the vicinity of the 
surface phonon frequency is determined by the surface phonon lifetime instead.
\\


In this work, we assume that $\tau_{sp}$ is constant, to be fitted to experiment.
Here, we discuss model description of the electron lifetime $\tau_e$.
Including relevant scattering mechanisms in our experiments, $\tau_e$ is given by,
\begin{align}
\tau_e(q,\omega) \approx \left[\tau_{0}^{-1}+\tau_{edge}(q)^{-1}+\tau_{ep}(\omega)^{-1}\right]^{-1}
\label{taue}
\end{align}
where $\tau_{0}$ describes a background damping due to scattering
with impurities and $\tau_{edge}(q)\approx a/(W-W_0)^{b}$ is related to scattering off
the ribbon edges. 
$\tau_{0}\approx 85\,$fs as measured from the Drude response of large area, unpatterned graphene. 
$a\approx 2\times 10^6$, of the order of Fermi velocity and
$b=1$ as discussed in the main text.
$\tau_{ep}(\omega)$ is electron lifetime due to
scattering with optical phonons.
It is related to the electron self-energy $\Sigma_{ep}$ via
$\tau_{ep}=\hbar/2\Im[\Sigma_{ep}]$ given by\cite{mahan00},
\begin{align}
\Sigma_{ep}(\omega)=-k_B T\sum_{\bold{q},\omega_{\lambda}}|M_{op}|^2 {\cal D}_{\lambda}(\omega_{\lambda})^0
{\cal G}^0(\bold{k}_f+\bold{q},\omega+\omega_{\lambda})
\end{align}
where ${\cal G}^0$ is the electron Green function and $\bold{k}_f$ is the Fermi wavevector.
According to density functional calculations, the
imaginary part of $\Sigma_{ep}$ can be approximated by\cite{PGCL07},
\begin{align}
\Im[\Sigma_{ep}(\omega)] = \gamma_0 \left|\hbar\omega+\hbar\omega_0+E_f\right|\times\frac{1}{2}
\left[\mbox{erf}\left(\frac{\hbar\omega-\hbar\omega_0}{\Delta_{ph}}\right)+\mbox{erf}\left(\frac{-\hbar\omega-\hbar\omega_0}{\Delta_{ph}}\right)+2\right]
\end{align}
where $\gamma_0$ describes the effective e-ph coupling and $\Delta_{ph}$
accounts for various energy broadening effects such as
the deviation from the Einstein phonon dispersion model.
They are estimated to be $\gamma_0\approx 0.018$ and 
$\Delta_{ph}\approx 50\,$meV from density 
function calculations\cite{PGCL07}.\\

As discussed previously, in the absence of interaction with
the surface phonons, the plasmon lifetime is simply $\tau_e(q,\omega=\omega_{pl})$.
In the presence of interaction with surface phonons,
the plasmon lifetime for the plasmon-phonon coupled modes
can be solved via Eq.\,\ref{plasphomodes} numerically,
with $\delta_{e}\approx [\tau_e(q,\omega=\omega_{pl}(q))]^{-1}$.
The computed plasmon
damping rates or lifetimes on DLC and SiO$_2$ substrates
are shown in Fig.\,4a and 4c of the main manuscript.
There, we assumed graphene doped at $E_f \sim -0.43\,$eV.\\

\begin{figure}[p]
\centering
\scalebox{0.6}[0.6]{\includegraphics*[viewport=0 20 640 730]{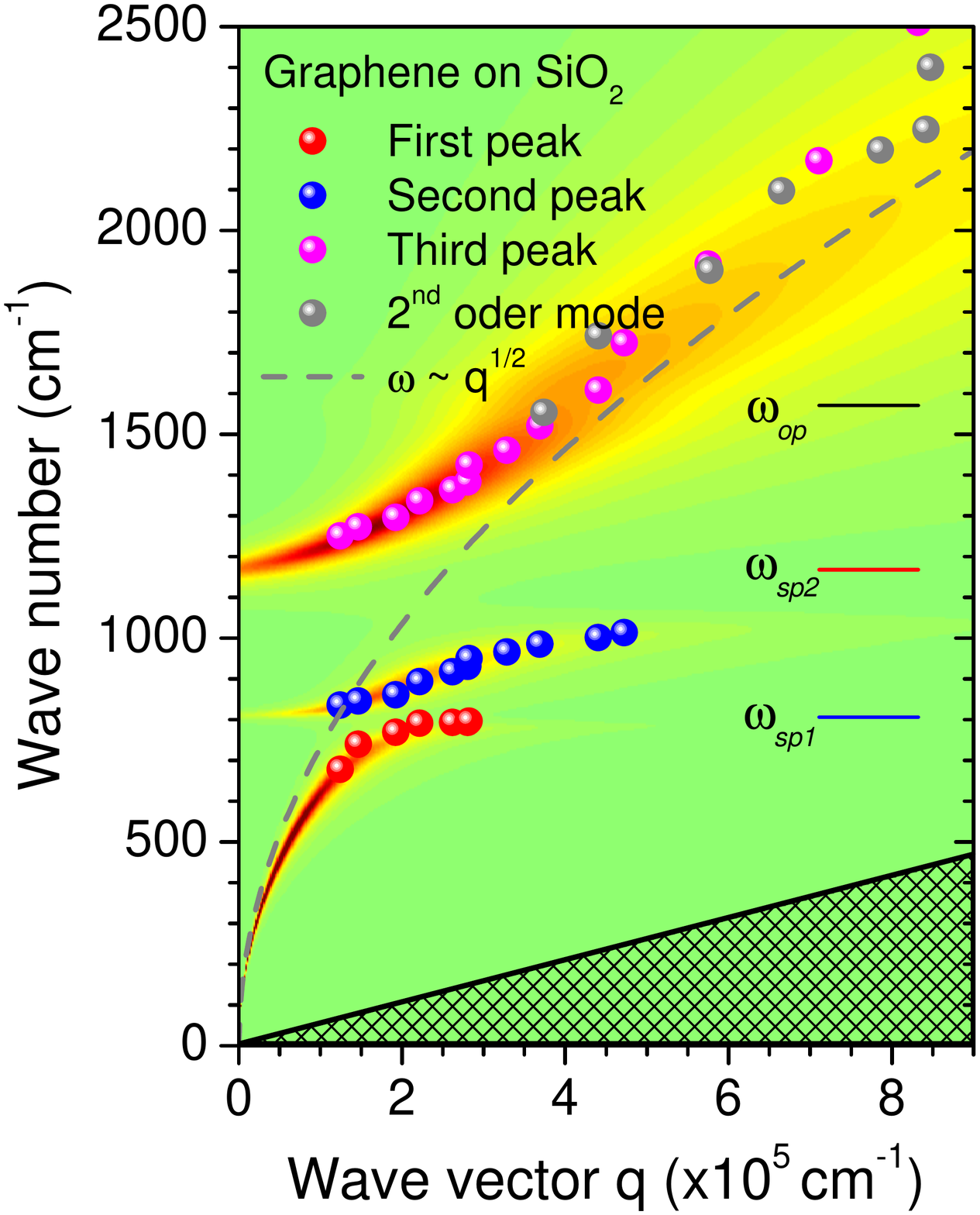}}
\caption{ \textbf{Graphene loss function on SiO$_2$ substrate.}
Calculated RPA loss function $\Im\left[1/\epsilon_T^{rpa}\right]$,
including interactions with the intrinsic and SiO$_2$ substrate phonons.
Graphene doping is assumed to be $E_f=-0.43\,$eV and an effective $\epsilon_{env}=1.5$.
Shaded regions represent the intraband Landau damping regime i.e. $\hbar\omega/E_f<q/k_F$.
Dashed line on the left plot is calculated from the classical
plasmon dispersion $\omega_{pl}^2=e^2 qv_F k_F/(2\pi\hbar\epsilon_0 \epsilon_{env})$.
The frequencies of the various phonon modes are assumed to be at
$\omega_{op}=1580\,$cm$^{-1}$, $\omega_{sp1}=806\,$cm$^{-1}$ and $\omega_{sp2}=1168\,$cm$^{-1}$.
The lifetime associated with the phonons used in these plots are
$\tau_{op}=70\,$fs, $\tau_{sp1}=0.5\,$ps and $\tau_{sp1}=0.2\,$ps.
Calculations include damping of single particle excitations
$\delta_e$ as described in the Suppl. info. text.
The coupling parameters used are $g_0=7.7\,$eV$\AA^{-1}$,
${\cal F}^2_{sp1}=0.2\,$meV and ${\cal F}^2_{sp2}=2\,$meV.
}
\label{fig1}
\end{figure}

Previously, we also show in Fig.\,3b of the main manuscript
the RPA loss function in SiO$_2$. However, the electron lifetime there
was simply assumed to be constant. In Fig.\,S\ref{fig1},
we calculate again the RPA loss function in SiO$_2$, 
but this time including the electron damping $\delta_e$ which describes 
our experiment as detailed in this section.
After the inclusion of a more accurate description of electron damping,
we note that the loss function can capture very well the plasmon peak intensity evolution 
for the three hybrid plasmon-phonon branches as seen experimentally, 
featuring two anti-crossings and spectral weight transfer 
from the low frequency to high frequency plasmon branch (peak 3) with increasing $q$.

\section{Resonance lineshape and the extraction of the plasmon damping }

This section describes the extraction of the plasmon damping rate from the
measured extinction spectra, used for Fig.\,4 of the main manuscript.\\

\begin{figure}[p]
\centering
\scalebox{0.55}[0.55]{\includegraphics*[viewport=0 0 750 730]{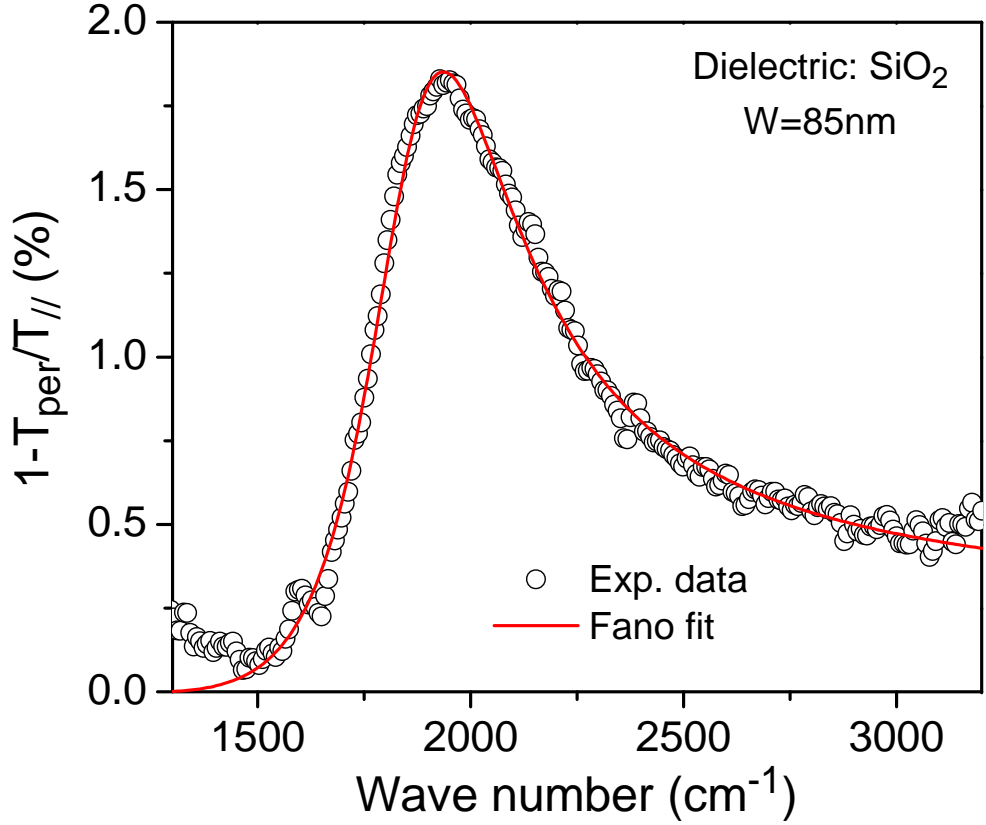}}
\caption{ \textbf{Resonance lineshape and fitting.}
Measured extinction spectra of a $W=85\,$nm
ribbon array on SiO$_2$ substrate, where we show only
part of the spectra relating to the peak 3 as described in the main manuscript.
The data can be fitted well by a Fano model, with
model and parameters described in Suppl. Info text.
}
\label{fig2}
\end{figure}

The far-IR plasmon resonace lineshape of graphene micro-disks and ribbons can be well 
described by a damped oscillator model\cite{YLC12},
which is derived from Drude conductivity. In the mid-IR regime, the lineshape becomes asymmetric, 
as seen for the third peak in Fig.\,3a of the main manuscript. 
Because the lineshape of the resonance peaks in the vicinity of the substrate phonon frequencies 
might be affected by the plasmon-phonon hybridization, here we focus on the third peak of 
relatively narrow ribbons on SiO$_2$ substrate, whose resonance frequency is far away from those 
of substrate phonons. Fig.\,S\ref{fig2} shows a spectrum (black curve) of a ribbon array with width 
$W=85\,$nm. The spectrum is very asymmetric and a Fano resonance model\cite{fano61,LLCB12} 
can well describe it, as shown by the red curve in the figure. 
In the Fano framework, the extinction spectrum is expressed as
\begin{align}
1-\frac{T_{per}}{T_{//}}=\frac{2p}{\pi \Gamma_p (1+q_f^2)}\frac{(q_f+\eta)^2}{(1+\eta^2)}
\end{align}
where $p$ is a parameter for the amplitude, 
$q_f$ is the Fano parameter, $\Gamma_p$ is the plasmon damping rate, 
and $\eta=2(\omega-\omega_0)/\Gamma_p$ with $\omega_0$
being the center frequency.
This equation is used to fit the spectra shown in the main manuscript to extract the 
plasmon damping rate $\Gamma_p$. Typical values for Fano parameter $q_f$ are around 3. \\

This Fano type resonance indicates that the plasmon resonance is interfered by a broad background continuum. 
As demonstrated before, the optical conductivity of graphene in the far-IR is Drude like. 
However, in the mid-IR range we are dealing with here, it's in the Pauli blocking regime, 
i.e., the optical conductivity has a very weak Drude response tail and some residue conductivity due to many-body effect. 
It is this residue conductivity serving as the broad continuum to form a Fano type resonance with the plasmon excitation.

\section{ Resonance broadening effect due to the ribbon Width inhomogeneity }

\begin{figure}[p]
\centering
\scalebox{0.55}[0.55]{\includegraphics*[viewport=0 0 750 730]{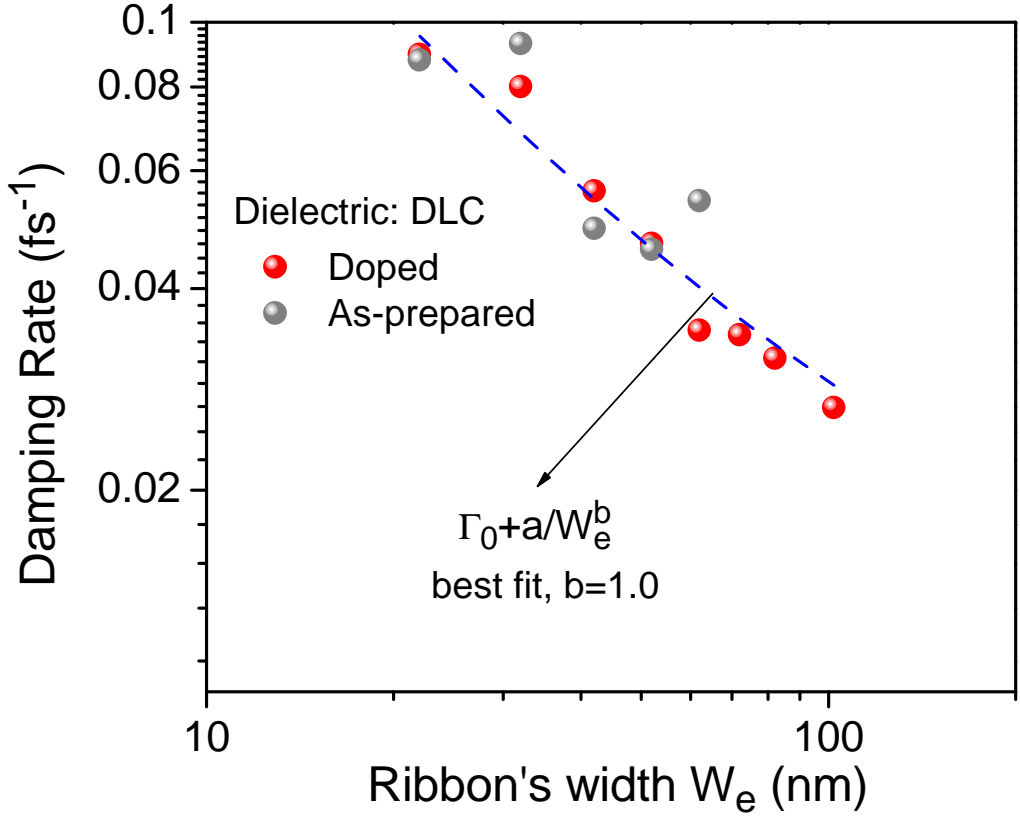}}
\caption{ \textbf{Damping rate scaling with width.}
Measured plasmon damping rate in graphene ribbons on 
DLC substrate as a function of the effective width $W_e=W-W_0$,
where $W$ is the physical width and $W_0=28\,nm$. See
also main text. 
The plasmon damping rate is best described by the 
scaling relation $\Gamma_0+a/W_e^b$, where $\Gamma_0=69cm^{-1}$
is related to the background damping, a quantity determined
from the Drude response of large area graphene. 
$a$ and $b$ are obtained from least-square-error fit to
the experimental data. $b$ was found to be $1.0$. 
}
\label{fig3}
\end{figure}

Long wavelength variations in the ribbon's width can also lead to an
apparent resonance broadening effect indistinguishable from lifetime broadening effects.
In DLC ribbons, the plasmon dispersion is described by a
simple $\omega_{pl}^2 = \alpha q$ dispersion, where $\alpha\equiv |E_f| e^2/2\pi\hbar^2\epsilon_0\epsilon_{env}$. 
In this case, the resonance broadening $\delta\omega_{pl}$ associated with a characteristic 
width inhomogeneity of $\Delta$ is given by,
\begin{align}
\delta\omega_{pl} \approx \frac{\sqrt{\alpha}\Delta}{2W_e^{\frac{3}{2}}}
\end{align}
In comparison to the plasmon lifetime broadening due to scattering off the edges,
it also has a power law behavior of the form $a/W_e^b$, the scaling exponent $b$ in this
case is $\tfrac{3}{2}$ instead of $1$.
Fig.\,S\ref{fig3} plots the experimentally extracted plasmon damping
as a function of the effective width $W_e$ for two different
doping levels. Least-square-error fit to
the data yields $b=1.0$, indicating that the broadening 
is due to carrier scattering off the edges. Furthermore,
as shown in Fig.\,S\ref{fig3},
there is also no noticeable dependence on doping, where
the width inhomogeneity effect would suggest otherwise.




\newpage

\renewcommand\figurename{FIG. S}

\end{document}